\documentclass[english]{revtex4}
\usepackage[T1]{fontenc}
\usepackage[latin9]{inputenc}
\usepackage{xcolor}
\usepackage{pdfcolmk}
\usepackage{amsmath}
\usepackage{amssymb}
\usepackage{esint}
\PassOptionsToPackage{normalem}{ulem}
\usepackage{ulem}

\makeatletter

\providecolor{lyxadded}{rgb}{0.99,0.0078,1}
\providecolor{lyxdeleted}{rgb}{1,0,0}

\DeclareRobustCommand{\lyxsout}[1]{\ifx\\#1\else\sout{#1}\fi}

\@ifundefined{textcolor}{}
{%
 \definecolor{BLACK}{gray}{0}
 \definecolor{WHITE}{gray}{1}
 \definecolor{RED}{rgb}{1,0,0}
 \definecolor{GREEN}{rgb}{0,1,0}
 \definecolor{BLUE}{rgb}{0,0,1}
 \definecolor{CYAN}{cmyk}{1,0,0,0}
 \definecolor{MAGENTA}{cmyk}{0,1,0,0}
 \definecolor{YELLOW}{cmyk}{0,0,1,0}
}

\usepackage{graphics}\usepackage{epstopdf}\usepackage{epsfig}\@ifundefined{definecolor}{\usepackage{color}}{}
\usepackage{amsfonts}\usepackage{bm}\usepackage{amscd}

\usepackage{makecell}

\usepackage{babel}

\usepackage{babel}

\makeatother

\usepackage{babel}
\begin{document}
\title{Nuclear interference by electronic de-orthogonalisation}
\author{Matisse Wei-Yuan Tu}
\affiliation{Max Planck Institute for the Structure and Dynamics of Matter, Luruper
Chaussee 149, 22761 Hamburg, Germany }
\author{Angel Rubio}
\affiliation{Max Planck Institute for the Structure and Dynamics of Matter, Luruper
Chaussee 149, 22761 Hamburg, Germany }
\author{E.K.U. Gross}
\affiliation{Quantum Dynamics Laboratory, Tsientang Institute of Advanced Study,
Hangzhou, China}
\affiliation{Fritz Haber Center for Molecular Dynamics, Institute of Chemistry,
The Hebrew University of Jerusalem, Jerusalem 91904 Israel}
\begin{abstract}
Interference is a universal consequence of superposition, yet in composite
quantum systems it can encode correlations between subsystems. We
show that in coupled electron--nuclear dynamics, interference in
the nuclear density can arise dynamically even when it is initially
absent. Starting from a superposition of orthogonal Born--Oppenheimer
electronic states, we demonstrate within the exact factorisation framework
that genuine non-adiabatic electron--nuclear correlations induce
de-orthogonalisation of the electronic factors, thereby generating
interference terms in the nuclear density. Such interference has no
counterpart in adiabatic evolution. Unlike conventional nuclear wave-packet
interference or interference that merely reflects electronic coherence
in a chosen basis, the effect identified here is a manifestation of
the compositeness of the full electron--nuclear state. Nuclear density
interference thus emerges as a direct dynamical signature of correlated
quantum motion in composite systems.
\end{abstract}
\maketitle

\section{Introduction}

The superposition principle is one of the corner stones of quantum
mechanics. Interference is a direct consequence of the superposition
principle. Whenever a wavefunction is written as a linear combination
of orthogonal component functions $\left\{ \Psi_{k}\left(x\right)\right\} $
of some dynamical variable $x$,

\begin{equation}
\Psi\left(x\right)=\sum_{k}c_{k}\Psi_{k}\left(x\right),\label{superpos-pr}
\end{equation}
its intensity separates into diagonal contributions and cross terms,
\begin{equation}
\left\vert \Psi\left(x\right)\right\vert ^{2}=\sum_{j}\left\vert c_{j}\right\vert ^{2}\left\vert \Psi_{j}\left(x\right)\right\vert ^{2}+\sum_{j\ne k}c_{j}^{*}c_{k}\Psi_{j}^{*}\left(x\right)\Psi_{k}\left(x\right).\label{intensity-supstr}
\end{equation}
The cross terms, proportional to $c_{j}^{*}c_{k}$, encode interference.
When the system under investigation is composite, consisting of two
or more coupled subsystems, interference at the level of a single
subsystem can be enriched or reshaped by correlations with the other
subsystems. The subsystem \textquotedblleft intensity\textquotedblright{}
is then no longer determined solely by superposition within that subsystem;
the cross terms may carry structural information about inter-subsystem
correlations, revealing features that would not arise in an isolated
system. A ubiquitous composite system is the coupled electron--nuclear
system, exemplified by molecules and solids. Here the two subsystems
are physically distinct: the light electrons and the much heavier
nuclei. Since nuclear motion organises the geometry of the system
and provides the spatial framework within which electronic states
are defined, interference manifested in the nuclear density is of
particular interest \cite{Zhang2022e2212114119,Han2023253202,Pedalino2026866}.

For electron-nuclear correlated systems, the joint dynamical variable
$x$ is given by $\left(\boldsymbol{r},\boldsymbol{R}\right)$, where
$\boldsymbol{r}$ and $\boldsymbol{R}$ are the electronic and the
nuclear coordinates respectively. The nuclear ($N$-body) density,
\begin{equation}
n\left(\boldsymbol{R},t\right)=\int\text{d}\boldsymbol{r}\left\vert \Psi\left(\boldsymbol{r},\boldsymbol{R},t\right)\right\vert ^{2}\label{nud-def}
\end{equation}
inherits the generic superposition structure,
\begin{equation}
n\left(\boldsymbol{R},t\right)=\sum_{j}\left\vert c_{j}\right\vert ^{2}n_{j}\left(\boldsymbol{R},t\right)+\sum_{j\ne k}c_{j}^{*}c_{k}n_{jk}\left(\boldsymbol{R},t\right),\label{nud-inf-itp}
\end{equation}
where
\begin{equation}
n_{j}\left(\boldsymbol{R},t\right)=\int\text{d}\boldsymbol{r}\left\vert \Psi_{j}\left(\boldsymbol{r},\boldsymbol{R},t\right)\right\vert ^{2},\label{nud-bkgrd}
\end{equation}
and

\begin{equation}
n_{jk}\left(\boldsymbol{R},t\right)=\int\text{d}\boldsymbol{r}\Psi_{j}^{*}\left(\boldsymbol{r},\boldsymbol{R},t\right)\Psi_{k}\left(\boldsymbol{r},\boldsymbol{R},t\right),\label{nud-inf-def}
\end{equation}
for all time $t$. The present study focuses on the interference encoded
in the off-diagonal contributions $n_{jk}\left(\boldsymbol{R},t\right)$
with $j\ne k$. In particular, we investigate how an initially vanishing
nuclear interference,
\begin{equation}
n_{jk}\left(\boldsymbol{R},t=0\right)=0,\label{init-no-nudinf-def}
\end{equation}
can dynamically become nonzero through the electron--nuclear correlations
embedded in each component of $\left\{ \Psi_{k}\left(\boldsymbol{r},\boldsymbol{R},t\right)\right\} $.
Nuclear interference phenomena in coupled electron--nuclear systems
have previously been analysed from perspectives that interference
either arises from superposition within a single nuclear channel \cite{Han2023253202,Chen2020033107}
or serves as a representation of electronic coherence in a particular
basis \cite{Zhang2022e2212114119}. In this context, the role of the
effective potential experienced by the nuclei has been emphasised
for the emergence of nuclear interference, since this potential directly
reflects electronic properties of the system. \cite{Babikov2005044315,Yuan20181289,Curchod2016034103}.
These studies together suggest that the manifestation of nuclear interference
is closely tied to electronic properties due to electron-nuclear compositeness.
In the present work, we focus on one particular aspect of such electronic
influence, namely how the electronic subsystem non-unitarity manifests
itself through the emergence of the nuclear interference contribution.
See Table \ref{wavIfcompare} for a comparison of various related
interference phenomena and more details in later discussion.

In the conventional Born--Huang (BH) representation of coupled electron--nuclear
dynamics, the full wavefunction is expanded as
\begin{equation}
\Psi\left(\boldsymbol{r},\boldsymbol{R},t\right)=\sum_{k}\chi_{k}\left(\boldsymbol{R},t\right)\left\langle \boldsymbol{r}\left\vert \varphi_{k}\left(\boldsymbol{R}\right)\right.\right\rangle ,\label{BH-def}
\end{equation}
where $\left\vert \varphi_{k}\left(\boldsymbol{R}\right)\right\rangle $
are Born--Oppenheimer (BO) electronic states. Owing to their orthogonality,
the nuclear density reduces to
\begin{equation}
n\left(\boldsymbol{R},t\right)=\sum_{k}\left\vert \chi_{k}\left(\boldsymbol{R},t\right)\right\vert ^{2},\label{nud-BH-g1}
\end{equation}
In this form the explicit cross-term structure proportional to $c_{j}^{*}c_{k}$
is no longer manifest; its dynamical emergence is concealed within
the coupled evolution of the nuclear amplitudes $\chi_{k}$, driven
by non-adiabatic couplings between BO electronic states. In principle,
one may solve the full time-dependent Schr\"{o}dinger equation (TDSE)
including these couplings and extract the nuclear interference from
the resulting dynamics \cite{Granucci1995228,Romstad199721,Donoso20007345,Curchod2016034103}.
However, full quantum treatments are computationally demanding \cite{BenNun20005161,Beck20001,Chen20061089,Yarkony2011481}.
Owing to the pronounced electron--nuclear mass disparity, many practical
approaches adopt mixed quantum--classical strategies in which nuclei
are propagated classically while electrons remain quantum mechanical.
This viewpoint underlies trajectory-based methods such as Ehrenfest
dynamics \cite{Alonso2008096403,Xavier09728}, surface hopping \cite{Tully901061,Tully98407,Schwartz965942,Prezhdo97825,Prezhdo975863,Hack019305,Bedard-Hearn05234106,Jasper05064103,Kaeb063197,Granucci10134111,Subotnik11024105,Subotnik16387,Shu23380},
and multi-trajectory implementations of exact factorisation (EF) \cite{Min15073001,Min20173048,Vindel-Zandbergen213852,EvaristoVillaseco2326380,Arribas2024233201}.

Within these frameworks, electronic decoherence induced by electron--nuclear
coupling has been an important research topic. While Ehrenfest dynamics
captures population transfer between BO states, it fails to describe
decoherence. Surface-hopping methods introduce explicit decoherence
corrections at the level of individual trajectories \cite{Tully901061,Tully98407,Schwartz965942,Prezhdo97825,Prezhdo975863,Hack019305,Bedard-Hearn05234106,Jasper05064103,Kaeb063197,Granucci10134111,Subotnik11024105,Subotnik16387,Shu23380}.
Multi-trajectory EF schemes generate decoherence and recoherence effects
intrinsically through inter-trajectory correlations. In this setting,
the so-called nuclear quantum momentum arising within EF has been
identified as a key mediator of electronic decoherence \cite{Min15073001,Min20173048,EvaristoVillaseco2326380,Arribas2024233201}.
Motivated by the long-standing interest in nuclear motion accompanying
inter-BO-state electronic transitions, we have previously demonstrated
within the EF framework that even along a single classical nuclear
trajectory, the non-unitary evolution of the electronic subsystem---rooted
in non-adiabatic electron--nuclear correlations---naturally gives
rise to electronic decoherence \cite{Tu2025043075}. Building on these
insights, we now turn to a complementary question: how does the subsystem
non-unitarity induced by electron--nuclear correlation manifest itself
in the interference contribution to the nuclear density?

To address this question, we adopt the exact factorisation (EF) framework
\cite{Abedi10123002,Suzuki14040501,Min15073001,Li2022113001,Arribas2024233201,Abedi1222A530,Min2014263004,Requist16042108,Requist17062503},
which provides separate equations of motion for the nuclear and electronic
subsystems while fully retaining their intrinsic coupling (a brief
recap is given in Appendix \ref{EF-EOM}). This formulation enables
a direct analysis of nuclear interference dynamics together with the
underlying electron--nuclear correlations and the associated electronic
evolution. We begin in Sec. \ref{main-trunk-inisAndNucd} by identifying
the mathematical condition required for the interference term to emerge
from an initially vanishing configuration, Eq. (\ref{init-no-nudinf-def}).
This condition, which we refer to as de-orthogonalisation, is manifested
in the electronic factors within EF. To elucidate its physical content,
we next consider dynamics initiated from a superposition of BO electronic
states, a representative case of orthogonal electronic components.
In Sec. \ref{main-trunk-Adia} we demonstrate that, under the adiabatic
approximation, the interference contribution remains identically zero,
thereby revealing the necessity of genuine non-adiabatic effects for
de-orthogonalisation to occur. We then analyse in Sec. \ref{main-trunk-NAND},
how the non-adiabatic correlation operator appearing in the EF electronic
equation of motion drives this de-orthogonalisation, thereby establishing
an explicit link between non-adiabaticity and the emergence of interference
in the nuclear density. Section \ref{main-trunk-example} illustrates
these mechanisms through fully correlated quantum-dynamical simulations
of a representative model system. Finally, Sec. \ref{main-trunk-otherPhenon}
situates our results within the broader landscape of related interference
phenomena and clarifies the rationale underlying Table \ref{wavIfcompare},
before we summarise our conclusions in Sec. \ref{conclu}.

\begin{table}[t] \caption{Fundamental relevance of compositeness and subsystem non-unitarity in representative interference phenomena.} 
\label{wavIfcompare} 
\centering 
\begin{tabular}{| c | c | c | } 
\hline
interference type & compositeness & subsystem non-unitarity \\ 
\hline
\makecell[l]{
(A): Interference within a single nuclear wave-packet 
}
& yes    
& no \\
\hline
\makecell[l]{
(B): Nuclear interference inferred from electronic coherence 
}
& yes & no  \\
\hline
\makecell[l]{
(C): Landau-Zener-St\"{u}ckelberg interference 
}
& no & no \\
\hline
\makecell[l]{
(D): Molecular matter-wave interference
}
& no & no \\
\hline
\makecell[l]{
(This work): Nuclear interference by electronic de-orthogonalisation
}
& yes & yes \\
\hline
\end{tabular} 
\end{table} 

\section{Results and discussions}

\label{main-trunk}

\subsection{Electronic de-orthogonalisation and nuclear interference}

\label{main-trunk-inisAndNucd}

Recall that if each wavefunction component $\Psi_{k}\left(\boldsymbol{r},\boldsymbol{R},t\right)$
satisfies the TDSE, then any superposition, e.g. Eq. (\ref{superpos-pr})
with time-independent coefficients $\left\{ c_{k}\right\} $, also
satisfies the same TDSE. Applying the exact factorisation to $\Psi_{k}\left(\boldsymbol{r},\boldsymbol{R},t\right)$
yields 
\begin{equation}
\Psi_{k}\left(\boldsymbol{r},\boldsymbol{R},t\right)=Y_{k}\left(\boldsymbol{R},t\right)\left\langle \boldsymbol{r}\left\vert \phi_{k}\left(t,\boldsymbol{R}\right)\right.\right\rangle ,\label{k_EF-def}
\end{equation}
where the nuclear factor $Y_{k}\left(\boldsymbol{R},t\right)$ is
normalised $\int\text{d}\boldsymbol{R}\left\vert Y_{k}\left(\boldsymbol{R},t\right)\right\vert ^{2}=1$
and the electronic factor $\left\vert \phi_{k}\left(t,\boldsymbol{R}\right)\right\rangle $
satisfies the partial normalisation condition $\left\langle \left.\phi_{k}\left(t,\boldsymbol{R}\right)\right\vert \phi_{k}\left(t,\boldsymbol{R}\right)\right\rangle =1$.
In terms of the EF factors, the nuclear density defined by Eq. (\ref{nud-def})
becomes
\begin{equation}
n\left(\boldsymbol{R},t\right)=\sum_{j}\left\vert c_{j}\right\vert ^{2}\left\vert Y_{j}\left(\boldsymbol{R},t\right)\right\vert ^{2}+\sum_{j\ne k}c_{j}^{*}c_{k}Y_{j}^{*}\left(\boldsymbol{R},t\right)Y_{k}\left(\boldsymbol{R},t\right)\left\langle \left.\phi_{j}\left(t,\boldsymbol{R}\right)\right\vert \phi_{k}\left(t,\boldsymbol{R}\right)\right\rangle .\label{nud-EF-gen}
\end{equation}
An initial no-interference state, Eq. (\ref{init-no-nudinf-def}),
is obtained by choosing orthogonal electronic factors at $t=0$,

\begin{equation}
\left\langle \left.\phi_{j}\left(t=0,\boldsymbol{R}\right)\right\vert \phi_{k}\left(t=0,\boldsymbol{R}\right)\right\rangle =\delta_{jk},\label{init-ortho-e-1}
\end{equation}
so that the nuclear density (defined by Eq. (\ref{nud-EF-gen})) reduces
to

\begin{equation}
n\left(\boldsymbol{R},t=0\right)=\sum_{j}\left\vert c_{j}\right\vert ^{2}\left\vert Y_{j}\left(\boldsymbol{R},t=0\right)\right\vert ^{2},\label{nud-EF-init}
\end{equation}
with 
\begin{equation}
\sum_{k}\left\vert c_{k}\right\vert ^{2}=1\label{init-coef-norm}
\end{equation}
ensured by normalisation.

\begin{figure}[h] \includegraphics[width=12cm, height=3.5 cm]{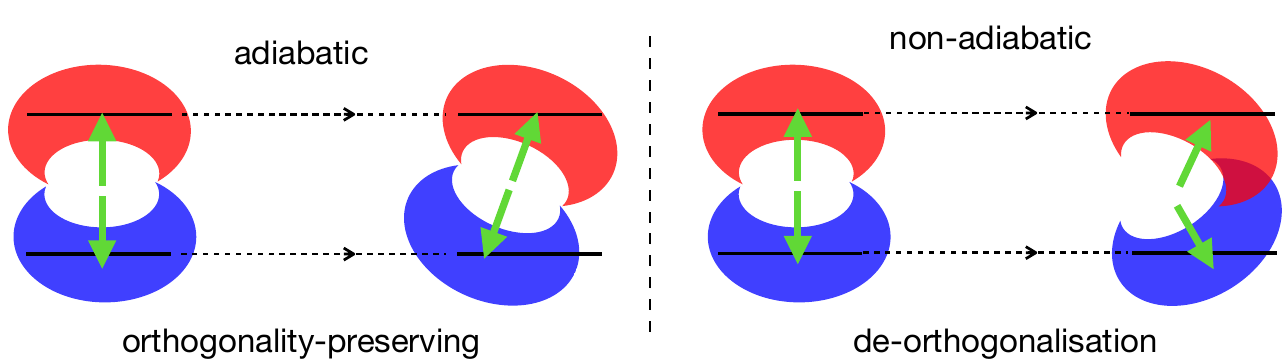} 
\caption{Illustration contrasting adiabatic evolution, which preserves the orthogonality of electronic states (left), with non-adiabatic evolution, which induces de-orthogonalisation (right).  Two initially orthogonal electronic wavefunctions are depicted as blue and red crescents with no overlap, and their orthogonality is further visualised by opposite green state vectors, in analogy with antipodal points on a Bloch sphere. Under adiabatic dynamics, the electronic states remain orthogonal throughout the evolution, yielding no interference contribution to the nuclear density.  Under non-adiabatic dynamics, orthogonality is not preserved, the electronic states acquire finite overlap, and an interference contribution emerges.  Solid black curves denote Born-Oppenheimer energy surfaces, while dashed arrows indicate the schematic direction of evolution.} 
\label{schematics} 
\end{figure}

In contrast to Eq. (\ref{nud-BH-g1}) where the dependence on the
coefficients $\left\{ c_{l}\right\} $ is implicit within the BH factors,
the EF-expression Eq. (\ref{nud-EF-gen}) explicitly displays the
interference contribution via the cross terms $c_{j}^{*}c_{k}$ with
$j\ne k$. If the initially orthogonal electronic factors later become
de-orthogonalised,
\begin{equation}
\left\langle \left.\phi_{j}\left(t,\boldsymbol{R}\right)\right\vert \phi_{k}\left(t,\boldsymbol{R}\right)\right\rangle \ne0,\label{jk-ovp-nonort-d1}
\end{equation}
for $j\ne k$, the nuclear density acquires nonzero interference contributions.
This electronic de-orthogonalisation-induced nuclear interference
is illustrated schematically in Fig. \ref{schematics}.

A natural starting point for an initial no-interference state is 
\begin{equation}
\Psi\left(\boldsymbol{r},\boldsymbol{R},t=0\right)=\chi_{ini}\left(\boldsymbol{R}\right)\sum_{k}c_{k}\left\langle \boldsymbol{r}\left\vert \varphi_{k}\left(\boldsymbol{R}\right)\right.\right\rangle ,\label{init-full-wvf-def}
\end{equation}
corresponding to the physically common situation in which the electronic
excitation is much faster than nuclear motion, so that the nuclei
are initially described by the same wavepacket while the electronic
state is prepared in a superposition of BO electronic states \cite{Granucci1995228,Arnold17033425,Jia194273,Golubev20083001,VillasecoArribas2024054102,FootNoteInitFullCor}.
Here $\chi_{ini}\left(\boldsymbol{R}\right)$ satisfies $\int\text{d}\boldsymbol{R}\left\vert \chi_{ini}\left(\boldsymbol{R}\right)\right\vert ^{2}=1$.
The initial value of each component is then 
\begin{equation}
\Psi_{k}\left(\boldsymbol{r},\boldsymbol{R},t=0\right)=\chi_{ini}\left(\boldsymbol{R}\right)\left\langle \boldsymbol{r}\left\vert \varphi_{k}\left(\boldsymbol{R}\right)\right.\right\rangle ,\label{k-it-gen}
\end{equation}
in accordance with Eq. (\ref{init-full-wvf-def}). Before explicitly
relating non-adiabatic effects to de-orthogonalisation, we first clarify
the outcome in their absence.

\subsection{The absence of nuclear density interference under the adiabatic processes}

\label{main-trunk-Adia}

From the definition Eq. (\ref{BH-def}) and the initial state Eq.
(\ref{init-full-wvf-def}), it is straightforward to see that
\begin{equation}
c_{k}\chi_{ini}\left(\boldsymbol{R}\right)=\chi_{k}\left(\boldsymbol{R},t=0\right).\label{init-nu-EF-BH-symb}
\end{equation}
The BO Hamiltonian $H^{BO}\left(\boldsymbol{R}\right)$ defines the
BO electronic eigenstates and eigenenergies as $H^{BO}\left(\boldsymbol{R}\right)\left\vert \varphi_{j}\left(\boldsymbol{R}\right)\right\rangle =\varepsilon_{j}\left(\boldsymbol{R}\right)\left\vert \varphi_{j}\left(\boldsymbol{R}\right)\right\rangle $.
Writing $\chi_{k}\left(\boldsymbol{R},t\right)=\left\langle \boldsymbol{R}\left\vert \chi_{k}\left(t\right)\right.\right\rangle $,
the adiabatic approximation to $\chi_{k}\left(\boldsymbol{R},t\right)\approx\chi_{k}^{ad}\left(\boldsymbol{R},t\right)=\left\langle \boldsymbol{R}\left\vert \chi_{k}^{ad}\left(t\right)\right.\right\rangle $
is obtained by setting the full wavefunction ansatz $\Psi_{k}\left(\boldsymbol{r},\boldsymbol{R},t\right)=\chi_{k}^{ad}\left(\boldsymbol{R},t\right)\left\langle \boldsymbol{r}\left\vert \varphi_{k}\left(\boldsymbol{R}\right)\right.\right\rangle $
to contain one and only one BO electronic state $\left\langle \boldsymbol{r}\left\vert \varphi_{k}\left(\boldsymbol{R}\right)\right.\right\rangle $.
Then through the variational principle leading to the familiar evolution
equation 
\begin{equation}
i\partial_{t}\left\vert \chi_{k}^{ad}\left(t\right)\right\rangle =H_{k}\left\vert \chi_{k}^{ad}\left(t\right)\right\rangle ,\label{ad-HBk-eq}
\end{equation}
where $H_{k}$ is the adiabatic nuclear Hamiltonian on the $k$th
BO potential energy surface (BOPES) specified by its scalar $\bar{\varepsilon}_{k}\left(\boldsymbol{R}\right)=\varepsilon_{k}\left(\boldsymbol{R}\right)+\mu\varepsilon^{NA}\left[\varphi_{k}\right]\left(\boldsymbol{R}\right)$
and vector $\boldsymbol{A}_{\nu}\left[\varphi_{k}\right]\left(\boldsymbol{R}\right)$
parts (see Eqs. (\ref{vecpt-def}) and (\ref{scapt-gi}) for the explicit
expressions of $\boldsymbol{A}_{\nu}\left[\varphi_{k}\right]\left(\boldsymbol{R}\right)$
and $\varepsilon^{NA}\left[\varphi_{k}\right]\left(\boldsymbol{R}\right)$.
Since $\left\vert \varphi_{k}\left(\boldsymbol{R}\right)\right\rangle $
is time-independent, these quantities carry no explicit time dependence.):
\begin{equation}
\left\langle \boldsymbol{R}\right\vert H_{k}\left\vert \chi\left(t\right)\right\rangle =\left\{ \mu\sum_{\nu}\frac{\left(-i\boldsymbol{\nabla}_{\nu}+\boldsymbol{A}_{\nu}\left[\varphi_{k}\right]\left(\boldsymbol{R}\right)\right)^{2}}{2M_{\nu}}+\bar{\varepsilon}_{k}\left(\boldsymbol{R}\right)\right\} \left\langle \boldsymbol{R}\left\vert \chi\left(t\right)\right.\right\rangle ,\label{ad-HBk-def}
\end{equation}
for any nuclear state $\left\vert \chi\left(t\right)\right\rangle $.
Atomic units are used and $\mu=m_{e}/M$ is the electronic-over-nuclear
mass ratio with a reference nuclear mass $M$ that $M_{\nu}$ is measured
in unit of $M$. Under this approximation, each component $\chi_{k}^{ad}\left(\boldsymbol{R},t\right)$
evolves independently on its own BOPES. Although the norm of $\left\vert \chi_{k}^{ad}\left(t\right)\right\rangle $
is preserved under Eq. (\ref{ad-HBk-eq}), it generally satisfies
$\left\langle \chi_{k}^{ad}\left(t\right)\left\vert \chi_{k}^{ad}\left(t\right)\right.\right\rangle =\left\vert c_{k}\right\vert ^{2}\ne1$.
By choosing $\left\langle \boldsymbol{R}\left\vert \chi\left(0\right)\right.\right\rangle =\chi_{ini}\left(\boldsymbol{R}\right)$
normalised to unity, the solution becomes $\left\vert \chi_{k}^{ad}\left(t\right)\right\rangle =c_{k}e^{-itH_{k}}\left\vert \chi\left(0\right)\right\rangle $.
Substituting this into Eq. (\ref{nud-BH-g1}) gives the nuclear density
\begin{equation}
n\left(\boldsymbol{R},t\right)=\sum_{k}\left\vert c_{k}\right\vert ^{2}\left\vert \left\langle \boldsymbol{R}\right\vert e^{-itH_{k}}\left\vert \chi\left(0\right)\right\rangle \right\vert ^{2},\label{nud-BH-ga}
\end{equation}
which clearly contains no interference contributions proportional
to $c_{j}^{*}c_{k}$ with $j\ne k$, in contrast to the general expression
Eq. (\ref{nud-inf-itp}). This absence of nuclear interference under
adiabatic dynamics, where only a single BO electronic state is retained
and hence no inter-BO-state transitions occur, can also be recovered
within the exact factorisation framework (see Appendix \ref{EF-EOM-ADA}).

\subsection{The rise of the nuclear density interference due to non-adiabatic
correlations}

\label{main-trunk-NAND}

We now incorporate non-adiabatic effects within the exact factorisation
(EF) framework. The initial condition for each component, Eq. (\ref{k-it-gen}),
translates into the EF factors as

\begin{equation}
\left\vert \phi_{k}\left(t=0,\boldsymbol{R}\right)\right\rangle =\left\vert \varphi_{k}\left(\boldsymbol{R}\right)\right\rangle ,\label{k_EF-el-it}
\end{equation}
and 
\begin{equation}
Y_{k}\left(\boldsymbol{R},t=0\right)=\chi_{ini}\left(\boldsymbol{R}\right).\label{k_EF-nuc-it}
\end{equation}
With this choice, the nuclear density at $t=0$ becomes $n\left(\boldsymbol{R},0\right)=\left\vert \chi_{ini}\left(\boldsymbol{R}\right)\right\vert ^{2}$
by virtue of Eq. (\ref{init-coef-norm}), and therefore contains no
initial interference contribution. 

The non-adiabatic electron-nuclear correlation manifests itself in
the EOM for the electronic factor in the EF framework via the electron-nuclear
correlation operator (ENC) $\mu V^{en}$ (see Eq. (\ref{e-EOM-set}))
that carries a small prefactor $\mu$, the electronic-over-nuclear
mass ratio. This suggests an approach to analyse the non-adiabatic
effects by perturbation expansion $\left\vert \phi_{k}\left(t,\boldsymbol{R}\right)\right\rangle =\left\vert \phi_{k}^{\left(0\right)}\left(t,\boldsymbol{R}\right)\right\rangle +\mu\left\vert \phi_{k}^{\left(1\right)}\left(t,\boldsymbol{R}\right)\right\rangle +\mathcal{O}\left(\mu^{2}\right)$
\cite{Scherrer17031035,Tu2025043075,Tu2025ARXIV251102004}. The zeroth
order without any non-adiabatic correlation simply follows $\left\vert \phi_{k}^{\left(0\right)}\left(t,\boldsymbol{R}\right)\right\rangle =e^{-i\varepsilon_{k}\left(\boldsymbol{R}\right)t}\left\vert \varphi_{k}\left(\boldsymbol{R}\right)\right\rangle $,
excluding any possibility of de-orthogonalisation, namely, the later-time
electronic states always stay orthogonal $\left\langle \left.\phi_{j}^{\left(0\right)}\left(t,\boldsymbol{R}\right)\right\vert \phi_{k}^{\left(0\right)}\left(t,\boldsymbol{R}\right)\right\rangle =0$
for $j\ne k$. The trivial case $\mu=0$, which automatically truncates
the perturbation series to only the zeroth order, indeed prevents
the nuclear degree of freedom from changing its density in time $\left\vert Y_{j}\left(\boldsymbol{R},t>0\right)\right\vert ^{2}=\left\vert Y_{j}\left(\boldsymbol{R},t=0\right)\right\vert ^{2}=\left\vert \chi_{ini}\left(\boldsymbol{R}\right)\right\vert ^{2}$
according to the EOM for the nuclear factor Eq. (\ref{EXF-EOM-nu}).
This trivial limit with no non-adiabatic transitions at all thus produces
no interference as expected.

The de-orthogonalisation required for interference therefore can only
rely on the non-adiabatic corrections, namely,
\begin{equation}
\left\langle \left.\phi_{j}\left(t,\boldsymbol{R}\right)\right\vert \phi_{k}\left(t,\boldsymbol{R}\right)\right\rangle =\mu S_{jk}^{\left(1\right)}\left(t,\boldsymbol{R}\right)+\mathcal{O}\left(\mu^{2}\right),\label{ovp-mu-expn-1}
\end{equation}
where $S_{jk}^{\left(1\right)}\left(t,\boldsymbol{R}\right)=\left\langle \left.\phi_{j}^{\left(1\right)}\left(t,\boldsymbol{R}\right)\right\vert \phi_{k}^{\left(0\right)}\left(t,\boldsymbol{R}\right)\right\rangle +\left\langle \left.\phi_{j}^{\left(0\right)}\left(t,\boldsymbol{R}\right)\right\vert \phi_{k}^{\left(1\right)}\left(t,\boldsymbol{R}\right)\right\rangle $
is the first-order correction to the overlapping $\left\langle \left.\phi_{j}\left(t,\boldsymbol{R}\right)\right\vert \phi_{k}\left(t,\boldsymbol{R}\right)\right\rangle $
in question. Carrying out the perturbation expansion to the first
order (detailed by Eqs. (\ref{1st-stat1}) and (\ref{1st-stat2})
), it is related to the ENC operator by
\begin{align}
 & S_{jk}^{\left(1\right)}\left(t,\boldsymbol{R}\right)\nonumber \\
= & i\int_{0}^{t}\text{d}t^{\prime}\left\{ \left[\left\langle \phi_{j}^{\left(0\right)}\left(t^{\prime},\boldsymbol{R}\right)\right\vert \left(V^{en}\left[\phi_{j}^{\left(0\right)},\left.\boldsymbol{\mathfrak{p}}\right\vert _{j}^{0}\right]\left(\boldsymbol{R},t^{\prime}\right)\right)^{\dagger}\right]\left\vert \phi_{k}^{\left(0\right)}\left(t^{\prime},\boldsymbol{R}\right)\right\rangle \right.\nonumber \\
 & \left.-\left\langle \phi_{j}^{\left(0\right)}\left(t^{\prime},\boldsymbol{R}\right)\right\vert \left[\text{\ensuremath{V^{en}\left[\phi_{k}^{\left(0\right)},\left.\boldsymbol{\mathfrak{p}}\right\vert _{k}^{0}\right]\left(\boldsymbol{R},t^{\prime}\right)\left\vert \phi_{k}^{\left(0\right)}\left(t^{\prime},\boldsymbol{R}\right)\right\rangle }}\right]\right\} ,\label{ovp-1st-1}
\end{align}
where $\left.\boldsymbol{\mathfrak{p}}\right\vert _{j}^{0}\left(\boldsymbol{R},t^{\prime}\right)$
is the nuclear momentum function (Eq. (\ref{pzero-q})) that one would
obtain if the electronic factor is initiated from the $j$th BO state.
That the result of Eq. (\ref{ovp-1st-1}) can be nonzero hinges on
the following properties of the ENC operator. First, $V^{en}$ (defined
by Eq. (\ref{Venmu})) itself depends on what it acts on. Explicitly
acting on the electronic factor $\phi_{k}^{\left(0\right)}$ but not
$\phi_{j}^{\left(0\right)}$ results in a $k$-dependent action also
through the appearance of the $k$-specific nuclear momentum function
$\left.\boldsymbol{\mathfrak{p}}\right\vert _{k}^{0}\left(\boldsymbol{R},t^{\prime}\right)$.
The difference between the nuclear response to different initial electronic
states contributes to $S_{jk}^{\left(1\right)}\left(t,\boldsymbol{R}\right)$.
Second, $V^{en}$ is non-hermitian (see related discussions in numerical
aspects \cite{Han20232186}) and physical implications for non-unitarity-associated
decoherence \cite{Tu2025043075}). For the usual perturbation caused
by a hermitian term added to the unperturbed Hamiltonian, neither
of the above two properties is satisfied. So two initially orthogonal
states will evolve to two later states that remain orthogonal to one
another in the ordinary case of a hermitian perturbation with orthogonality-preserving
unitary dynamics. The non-adiabatic perturbtaion is thus characterised
by a fundamentally different mathematical structure, namely, de-orthogonalisation,
that can have a consequence on physical behaviour, e.g., the interference
contribution to the nuclear density. 

\subsection{example}

\label{main-trunk-example}

We now numerically corroborate the analytically established connection
between electronic de-orthogonalisation, non-adiabaticity, and the
resulting interference contributions to the nuclear density by solving
the full quantum dynamics of a simple but representative model system.
Our goal is to explicitly resolve how de-orthogonalisation and nuclear-density
interference align across different regions of nuclear configuration
space.

\begin{figure}[h] \includegraphics[width=12cm, height=4.5 cm]{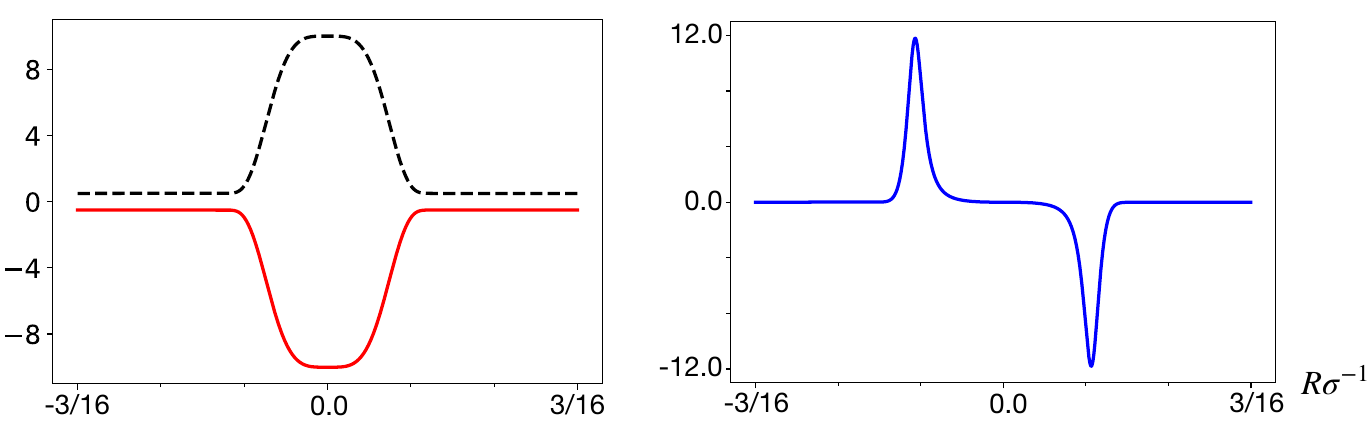} 
\caption{BO energy landscape (left, in unit of $g_{0}$) and the NAC profile (right, in unit of $g_{0}/\sigma$) whose quantum dynamics is numerically studied here as an example.  The NAC profile shows two distinct regions strong couplings separated by a region of larger BO gap. } 
\label{BOlandscape} 
\end{figure}

To realise multiple well-separated non-adiabatic regions in the simplest
setting, we adopt the widely used double-arc model \cite{Subotnik11024105,Curchod2013184112,Agostini162127,Ha22174109}.
This is a one-dimensional ($\boldsymbol{R}\rightarrow R$) two-state
model defined through the Born--Oppenheimer Hamiltonian $H^{BO}\left(R\right)=V_{0}\left(R\right)+\frac{g_{0}}{2}\sigma_{z}+V_{x}\left(R\right)\sigma_{x}$,
where $\sigma_{i}$ ($i=x,y,z$) are Pauli matrices and the constant
$g_{0}>0$ sets the characteristic electronic energy scale. In this
construction, the strength and localisation of the non-adiabatic couplings
(NACs) are controlled entirely by the $R$-dependence of $V_{x}\left(R\right)$,
whereas $V_{0}\left(R\right)$ provides only a uniform shift of the
BO energies and does not affect the BO eigenstates $\left\{ \left\vert \varphi_{0}\left(R\right)\right\rangle ,\left\vert \varphi_{1}\left(R\right)\right\rangle \right\} $
or the associated NACs. More generally, since first-order NACs scale
inversely with the BO energy gap, strong non-adiabatic effects are
expected to emerge in regions where the gap becomes small. The double-arc
model is designed precisely to capture this structure, producing two
localised regions of strong coupling separated by an intermediate
region of large BO splitting \cite{Subotnik11024105,Curchod2013184112,Agostini162127,Ha22174109}.
Here we implement this behaviour smoothly by choosing $V_{x}\left(R\right)=g_{x}\exp\left(-\kappa\left(R/L_{x}\right)^{\alpha}\right)$
and $V_{0}\left(R\right)=K\left(R/L_{W}\right)^{2}$, with $K$ and
$g_{x}$ measured in units of $g_{0}$, and $L_{W}$ and $L_{x}$
measured in units of the initial nuclear wavepacket width $\sigma$,
namely, $\chi_{ini}\left(R\right)=\left[\sigma\sqrt{\pi}\right]^{-1/2}\exp\left[-R^{2}/\left(2\sigma^{2}\right)\right]$.
Throughout this illustrative discussion we fix $L_{W}/\sigma=15$,
$K/g_{0}=0.1$, $\alpha=4$, and $g_{x}/g_{0}=10$, yielding pronounced
non-adiabatic regions. The resulting BO energy surfaces and NAC profile
are shown in Fig. \ref{BOlandscape}. The nuclear dynamics is solved
on a discretised real-space grid, where the kinetic energy is represented
by nearest-neighbour hopping with amplitude $J_{L}$. With the above
parameters fixed, the character of the non-adiabatic evolution is
primarily controlled by $J_{L}$, which sets the nuclear kinetic scale
common to electron--nuclear correlated models, as well influenced
by the model-specific parameter $\kappa$.

\begin{figure}[h] \includegraphics[width=12cm, height=8.0 cm]{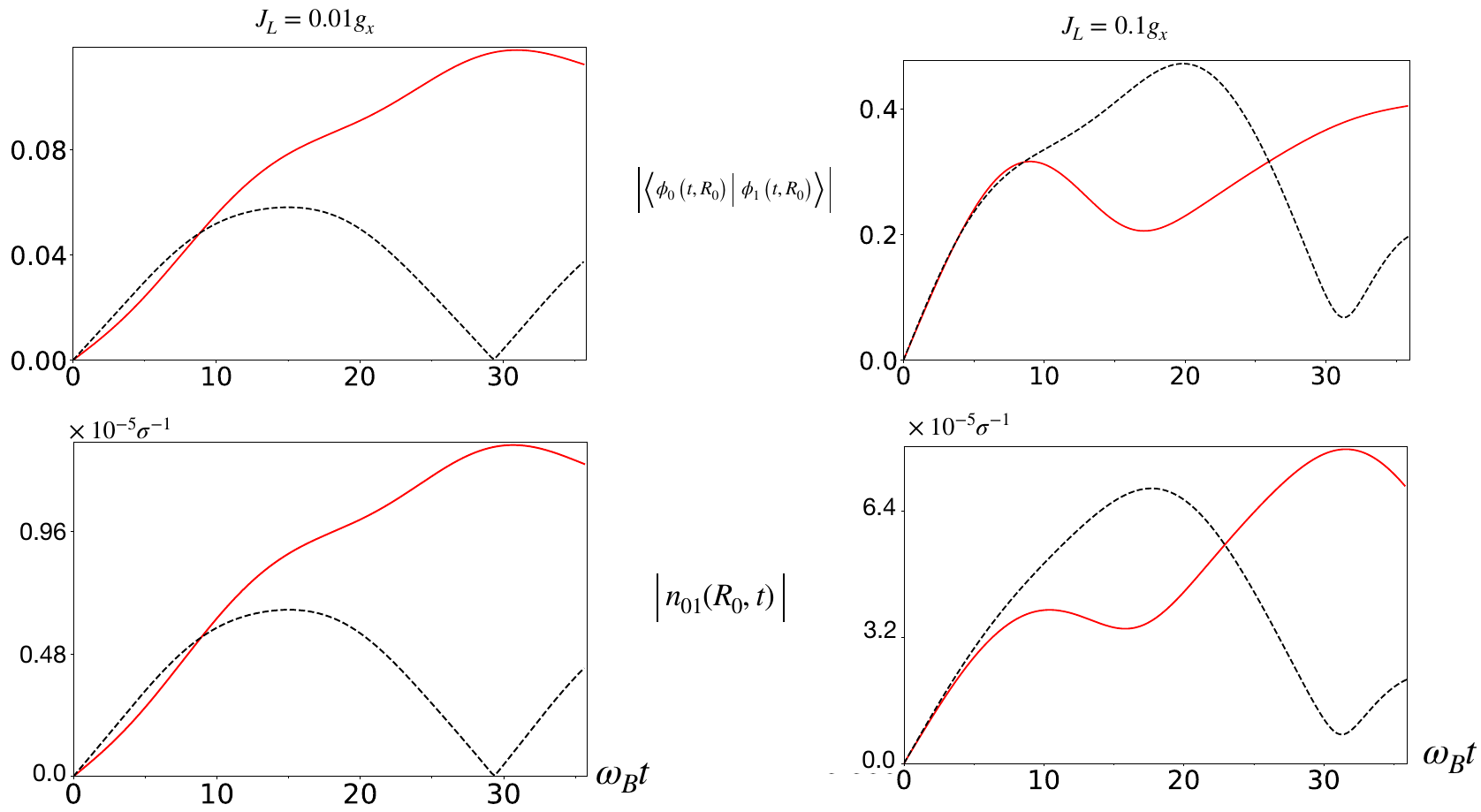} 
\caption{The concurrent growth of electronic de-orthogonalisation (upper rows) and nuclear interference (lower rows) under weaker (left column) and stronger (right column) non-adiabatic conditions represented by $J_{L}$ (see legends).  The red solid/black dashed curves are all with $\kappa=1.0$/$\kappa=2.5$.  For each $(J_{L},\kappa)$, $R_{0}$ is chosen as the coordinate at which  $\left|\left\langle \phi_{0}(t,R)\middle|\phi_{1}(t,R)\right\rangle\right|$ reaches its maximum within the simulation time window.  For the left column, the red solid/black dashed curves are with $R_{0}=0.075\sigma$/$R_{0}=0.05\sigma$. For the right column,  the red solid/black dashed curves are with $R_{0}=0.1\sigma$/$R_{0}=0.075\sigma$. Here $\omega_{B}t$ presents time as a dimensionless parameter in which we use $\omega_{B}=19.25g_{0}$/$\omega_{B}=20.15g_{0}$ for the left/right column as the estimated energy width of the fully correlated system.} 
\label{de-ort-evo} 
\end{figure}

The numerical calculations also confirm that $n_{01}\left(R,t\right)$
vanishes either under the setting of $J_{L}=0$ (corresponding to
$\mu=0$ in the previous discussions) or $g_{x}=0$ (i.e., in the
absence of NACs), irrespective of the other parameter choices. The
degree of de-orthogonalisation is quantified by $\left\vert \left\langle \left.\phi_{0}\left(t,R\right)\right\vert \phi_{1}\left(t,R\right)\right\rangle \right\vert $
which is initially zero for all $R$ and is bounded above by unity.
Figure \ref{de-ort-evo} illustrates how this overlap (upper row)
grows from zero to finite values in time at a representative coordinate
$R=R_{0}$ (see caption). The corresponding rise of the interference
contribution $\left\vert n_{01}\left(R,t\right)\right\vert $ is shown
in the lower row. As expected, larger $J_{L}$, corresponding to stronger
non-adiabatic effects, leads to more pronounced de-orthogonalisation
and hence larger interference magnitudes. We also compare different
values of $\kappa$ at fixed $J_{L}$ (see the two curves in each
column of Fig. \ref{de-ort-evo}. The early-time ordering of these
curves differs between the weakly and strongly non-adiabatic regimes,
reflecting the fact that $\kappa$ is a model-specific parameter,
whereas $J_{L}$ sets the universal nuclear kinetic scale. Nevertheless,
for all parameter choices, both $\left\vert \left\langle \left.\phi_{0}\left(t,R\right)\right\vert \phi_{1}\left(t,R\right)\right\rangle \right\vert $
and $\left\vert n_{01}\left(R,t\right)\right\vert $ exhibit a clear
monotonic growth over an initial onset period, consistent with the
general analytical picture developed above.

\begin{figure}[h] \includegraphics[width=12cm, height=7.0 cm]{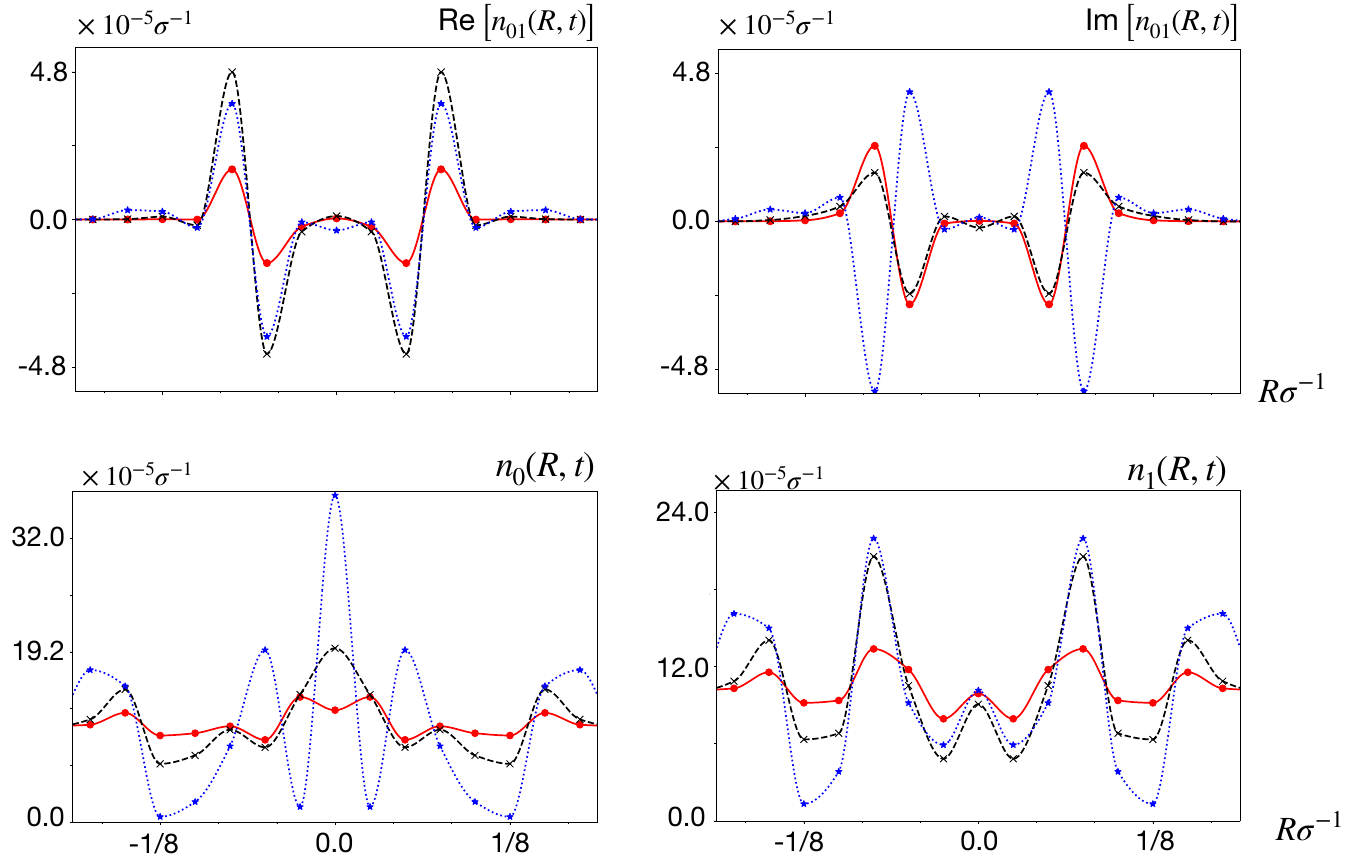} 
\caption{Snapshots of the profiles for the interference contribution $n_{01}(R,t)$ (upper panel with the real/imaginary part on the left/right column) and for the non-interference contributions $n_{0}(R,t)$ and $n_{1}(R,t)$ (lower panel) to the nuclear density for $J_{L}=0.1\,g_{x}$ and $\kappa=2.5$ (parameters used by the black dashed curves in the right column of Fig. \ref{de-ort-evo}).  Discrete data points at $\omega_{B}t=5,\,10,\,20$ are shown as red circles, black crosses, and blue stars, respectively. Smooth solid, dashed, and dotted curves are drawn through the data as guides to the eye. } 
\label{infamp-snapshots} 
\end{figure}

In Fig. \ref{infamp-snapshots}, we present snapshots of the $R$-dependent
profiles of the different contributions to the nuclear density, taken
during the time interval in which de-orthogonalisation grows monotonically.
The upper row shows the interference term $n_{01}\left(R,t\right)$,
plotted separately in its real (left) and imaginary (right) parts.
Both components can contribute to the interference term $\sum_{j\ne k}c_{j}^{*}c_{k}n_{jk}\left(R,t\right)=\left\vert c_{0}\right\vert \left\vert c_{1}\right\vert \left[\cos\phi\text{Re}\left(n_{01}\left(R,t\right)\right)-\sin\phi\text{Im}\left(n_{01}\left(R,t\right)\right)\right]$,
where $\phi$ is the relative phase between the initial superposition
coefficients $c_{0}$ and $c_{1}$. The lower row shows the other
contributions $n_{0}\left(R,t\right)$ and $n_{1}\left(R,t\right)$.
To connect these profiles with non-adiabaticity, we compare their
$R$-dependence with that of the NAC shown in the lower right panel
of Fig. \ref{schematics}. Notably, $n_{01}\left(R,t\right)$ closely
follows the NAC profile, becoming significant only in the regions
of enhanced non-adiabatic coupling. By contrast, the profiles of $n_{0}\left(R,t\right)$
and $n_{1}\left(R,t\right)$ do not display an obvious correlation
with the NAC. This example therefore confirms that the emergence of
interference in the nuclear density is intimately tied to the non-adiabatic
coupling and the underlying electronic de-orthogonalisation. 

\subsection{Comparison to other related phenomena}

\label{main-trunk-otherPhenon}

\subsubsection{electronic decoherence between pairs of BO states}

In our previous work, we showed that non-adiabatic electron--nuclear
correlations render the evolution of the electronic factor non-unitary,
thereby inducing electronic decoherence even along a single classical
nuclear trajectory \cite{Tu2025043075}. It is therefore essential
to distinguish two distinct consequences of this non-unitarity: decoherence
and de-orthogonalisation.

Electronic decoherence refers to the decay of coherence defined pairwise
with respect to two prescribed BO states, $\varphi_{l}$ and $\varphi_{m}$
with $l\ne m$ and is quantified by the off-diagonal elements of the
reduced electronic density matrix, $\rho_{lm}^{e}\left(t\right)\equiv\int\text{d}\boldsymbol{R}\int\text{d}\boldsymbol{r}\left\langle \varphi_{l}\left(\boldsymbol{R}\right)\left\vert \boldsymbol{r}\right.\right\rangle \left\vert \Psi\left(\boldsymbol{r},\boldsymbol{R},t\right)\right\vert ^{2}\left\langle \boldsymbol{r}\left\vert \varphi_{m}\left(\boldsymbol{R}\right)\right.\right\rangle $
\cite{VillasecoArribas2024054102,Arribas2024233201}. Expressed in
terms of the EF factors (see Eq. (\ref{k_EF-def})), this quantity
decomposes as $\rho_{lm}^{e}\left(t\right)=\sum_{j}\left\vert c_{j}\right\vert ^{2}\rho_{lm}^{j}\left(t\right)+\sum_{j\ne k}c_{j}^{*}c_{k}\rho_{lm}^{jk}\left(t\right)$,
with $\rho_{lm}^{j}\left(t\right)=\int\text{d}\boldsymbol{R}\left\vert Y_{j}\left(\boldsymbol{R},t\right)\right\vert ^{2}\left\langle \left.\phi_{j}\left(t,\boldsymbol{R}\right)\right\vert \varphi_{m}\left(\boldsymbol{R}\right)\right\rangle \left\langle \left.\varphi_{l}\left(\boldsymbol{R}\right)\right\vert \phi_{j}\left(t,\boldsymbol{R}\right)\right\rangle $
and $\rho_{lm}^{jk}\left(t\right)=\int\text{d}\boldsymbol{R}Y_{j}^{*}\left(\boldsymbol{R},t\right)Y_{k}\left(\boldsymbol{R},t\right)\left\langle \left.\phi_{j}\left(t,\boldsymbol{R}\right)\right\vert \varphi_{m}\left(\boldsymbol{R}\right)\right\rangle \left\langle \left.\varphi_{l}\left(\boldsymbol{R}\right)\right\vert \phi_{k}\left(t,\boldsymbol{R}\right)\right\rangle $.
Crucially, the coherence by definition involves products of two projection
amplitudes onto prescribed BO states, namely, $\left\langle \left.\phi_{j}\left(t,\boldsymbol{R}\right)\right\vert \varphi_{m}\left(\boldsymbol{R}\right)\right\rangle $
and $\left\langle \left.\varphi_{l}\left(\boldsymbol{R}\right)\right\vert \phi_{k}\left(t,\boldsymbol{R}\right)\right\rangle $,
rather than a single overlap between conditional electronic states.
Under the adiabatic approximation, the coherence is reduced to $\rho_{lm}^{e}\left(t\right)=c_{m}^{*}c_{l}\int\text{d}\boldsymbol{R}Y_{m}^{*}\left(\boldsymbol{R},t\right)Y_{l}\left(\boldsymbol{R},t\right)$
governed only by the overlap between the nuclear factors. By contrast,
the de-orthogonalisation discussed in the present work is governed
directly by the electronic-factor overlap $\left\langle \left.\phi_{j}\left(t,\boldsymbol{R}\right)\right\vert \phi_{k}\left(t,\boldsymbol{R}\right)\right\rangle $
itself. Although both effects originate from the non-unitarity of
the electronic evolution, de-orthogonalisation constitutes a physically
distinct manifestation of non-adiabatic dynamics. Its consequences
extend beyond electronic decoherence and appear explicitly in the
nuclear density.

\subsubsection{other interference phenomena}

Here we compare the nuclear interference induced by non-adiabatic
de-orthogonalisation identified in this work to several related interference
phenomena listed in Table \ref{wavIfcompare}. Types (A) and (B) are
closest in spirit, since they concern interference effects carried
by the nuclear degree of freedom within electron--nuclear composite
systems. By contrast, (C) Landau--Zener--St\"{u}ckelberg interference
\cite{Shevchenko20101} is a driven two-level phenomenon, where non-adiabaticity
is imposed by an external classical driving, fundamentally distinct
from the intrinsic non-adiabatic electron--nuclear correlations responsible
for de-orthogonalisation here. Finally, (D) molecular matter-wave
interference \cite{Hornberger2012157,Pedalino2026866} represents
the spatial interference where the molecule as a whole acts as the
quantum mechanical particle. Electronic overlap may modulate nuclear
patterns but without invoking the adiabatic versus non-adiabatic correlation
central to the present mechanism. We therefore focus below on clarifying
the distinctions between (A) and (B) and the interference regime addressed
here.

To place the discussion in context, recall the familiar mechanism
by which interference arises in a single quantum system. Consider
an initial superposition of two orthogonal states, 
\begin{equation}
\left\vert \Psi\left(t=0\right)\right\rangle =c_{0}\left\vert \psi_{0}\right\rangle +c_{1}\left\vert \psi_{1}\right\rangle .\label{1qe-init-demo}
\end{equation}
If the Hamiltonian dynamically couples these states, each initial
component $\left\vert \Psi_{k}\left(t=0\right)\right\rangle =\left\vert \psi_{k}\right\rangle $
develops amplitudes in both channels, namely,
\begin{equation}
\left\vert \Psi_{k}\left(t\right)\right\rangle =\sum_{l}\xi_{l\leftarrow k}\left(t\right)\left\vert \psi_{l}\right\rangle ,\label{1qe-psi-k}
\end{equation}
where $\xi_{l\leftarrow k}\left(t=0\right)=\delta_{lk}$. The intensity
of the later-state $\left\vert \Psi\left(t\right)\right\rangle =c_{0}\left\vert \Psi_{0}\left(t\right)\right\rangle +c_{1}\left\vert \Psi_{1}\left(t\right)\right\rangle $
then contains terms proportional to
\begin{equation}
c_{1}^{*}c_{0}\xi_{l\leftarrow1}^{*}\left(t\right)\xi_{l\leftarrow0}\left(t\right),\label{1qe-psiq-inf}
\end{equation}
which vanish initially but become nonzero through dynamical amplitude
transfer between the orthogonal channels. In this sense, interference
is commonly understood as arising when amplitudes originating from
distinct orthogonal states are brought into the same channel by the
dynamics.

Consider an arbitrary orthonormal electronic basis $\left\{ \left\vert \psi_{k}\left(\boldsymbol{R}\right)\right\rangle \right\} $
which may be chosen as a diabatic basis or a BO basis. Let the initial
component states be
\begin{equation}
\Psi_{k}\left(\boldsymbol{r},\boldsymbol{R},0\right)=\zeta_{k}\left(\boldsymbol{R}\right)\left\langle \boldsymbol{r}\left\vert \psi_{k}\left(\boldsymbol{R}\right)\right.\right\rangle ,\label{init-k-gen}
\end{equation}
with arbitrary normalised function $\zeta_{k}\left(\boldsymbol{R}\right)$.
The full wavefunction is then $\Psi\left(\boldsymbol{r},\boldsymbol{R},t\right)=\sum_{k}c_{k}\Psi_{k}\left(\boldsymbol{r},\boldsymbol{R},t\right)$
where each $\Psi_{k}\left(\boldsymbol{r},\boldsymbol{R},t\right)$
evolves from its corresponding initial component Eq. (\ref{init-k-gen}).
Expanding the time-evolved state in the same electronic basis gives
\begin{equation}
\Psi_{k}\left(\boldsymbol{r},\boldsymbol{R},t\right)=\sum_{l}\xi_{l\leftarrow k}\left(\boldsymbol{R},t\right)\left\langle \boldsymbol{r}\left\vert \psi_{l}\left(\boldsymbol{R}\right)\right.\right\rangle ,\label{psi-k-t-arb-basis-1}
\end{equation}
The coefficients $\xi_{l\leftarrow k}\left(\boldsymbol{R},t\right)$
describe amplitude transfer between electronic channels, with $l\leftarrow k$
indicating that amplitude in channel $l$ originated from the initial
channel $k$. The nuclear density is then written as
\begin{equation}
n\left(\boldsymbol{R},t\right)=\sum_{l}\left\vert \Xi_{l}\left(\boldsymbol{R},t;\left\{ c\right\} \right)\right\vert ^{2},\label{psi-t-arb-basis-n1}
\end{equation}
where 
\begin{equation}
\Xi_{l}\left(\boldsymbol{R},t;\left\{ c\right\} \right)=\sum_{k}c_{k}\xi_{l\leftarrow k}\left(\boldsymbol{R},t\right).\label{psi-t-arb-basis-s1}
\end{equation}
From this representation, the interference contribution associated
with $c_{j}^{*}c_{k}$ takes the form
\begin{equation}
n_{jk}\left(\boldsymbol{R},t\right)=\sum_{l}\xi_{l\leftarrow j}^{*}\left(\boldsymbol{R},t\right)\xi_{l\leftarrow k}\left(\boldsymbol{R},t\right).\label{psi-k-psiq-inf}
\end{equation}
This is the familiar amplitude-transfer picture of interference. Nuclear
interference appears because dynamics generates non-vanishing amplitudes
$\xi_{l\leftarrow j}\left(\boldsymbol{R},t\right)$ and $\xi_{l\leftarrow k}\left(\boldsymbol{R},t\right)$
within common electronic channel $l$.

Within this representation, the commonly discussed types (A) and (B)
of nuclear interference can be understood directly. Type (B) interference
refers to interference between nuclear wave packets associated with
different electronic channels. Defining the electronic coherence $\rho_{ll^{\prime}}(t)=\int\mathrm{d}\boldsymbol{R}\Xi_{l^{\prime}}^{*}\left(\boldsymbol{R},t;\left\{ c\right\} \right)\Xi_{l}\left(\boldsymbol{R},t;\left\{ c\right\} \right)$
the cross term $\Xi_{l^{\prime}}^{*}\left(\boldsymbol{R},t;\left\{ c\right\} \right)\Xi_{l}\left(\boldsymbol{R},t;\left\{ c\right\} \right)$
for $l\neq l'$ is interpreted as the interference \cite{Zhang2022e2212114119}.
Type (A) interference instead focuses on a single nuclear wave packet
$\Xi_{l}\left(\boldsymbol{R},t;\left\{ c\right\} \right)$ \cite{Babikov2005044315,Yuan20181289}.
If this wave packet is itself decomposed into several pathway contributions,
$\Xi_{l}\left(\boldsymbol{R},t;\left\{ c\right\} \right)=\sum_{\alpha}A_{\alpha}^{\left(l\right)}\left(\boldsymbol{R},t\right)$,
then interference can also arise from cross terms $A_{\alpha}^{\left(l\right)*}\left(\boldsymbol{R},t\right)A_{\alpha^{\prime}}^{\left(l\right)}\left(\boldsymbol{R},t\right)$
with $\alpha\ne\alpha^{\prime}$ \cite{Granucci1995228,Chen2020033107,Han2023253202}.
Hence both type (A) and type (B) analyse interference through overlaps
between amplitudes appearing in nuclear wave packets, differing only
in whether the interfering amplitudes belong to the same or to different
electronic channels.

Applying exact factorisation separately to each component $\Psi_{k}\left(\boldsymbol{r},\boldsymbol{R},t\right)$
using Eq. (\ref{k_EF-def}) yields the equivalent relation
\begin{equation}
n_{jk}\left(\boldsymbol{R},t\right)=\sum_{l}\xi_{l\leftarrow j}^{*}\left(\boldsymbol{R},t\right)\xi_{l\leftarrow k}\left(\boldsymbol{R},t\right)=Y_{j}^{*}\left(\boldsymbol{R},t\right)Y_{k}\left(\boldsymbol{R},t\right)\left\langle \left.\phi_{j}\left(t,\boldsymbol{R}\right)\right\vert \phi_{k}\left(t,\boldsymbol{R}\right)\right\rangle .\label{arb-basis-nucd-cross}
\end{equation}
This identity establishes the connection between the two viewpoints.
The left-hand side expressed through $\xi_{l\leftarrow k}\left(\boldsymbol{R},t\right)$
emphasises amplitude transfer between electronic channels, whereas
the exact-factorisation form reveals that the same interference contribution
can be viewed as arising from the overlap of conditional electronic
states. Initially, $\left\langle \left.\phi_{j}\left(t=0,\boldsymbol{R}\right)\right\vert \phi_{k}\left(t=0,\boldsymbol{R}\right)\right\rangle =\left\langle \psi_{j}\left(\boldsymbol{R}\right)\left\vert \psi_{k}\left(\boldsymbol{R}\right)\right.\right\rangle =0$
for $j\ne k$. Therefore $n_{jk}\left(\boldsymbol{R},0\right)=0$.
A non-zero interference contribution requires $\left\langle \left.\phi_{j}\left(t,\boldsymbol{R}\right)\right\vert \phi_{k}\left(t,\boldsymbol{R}\right)\right\rangle \ne0$,
namely electronic de-orthogonalisation. Unlike type (A), which focuses
on interference within a single nuclear wave packet, or type (B),
which focuses on coherence between pairs of nuclear wave packets associated
with different electronic channels, the present mechanism identifies
a subsystem property of the electrons themselves---loss of orthogonality
of conditional electronic states---as the quantity enabling the interference
term in the nuclear density.

Importantly, the appearance of the electronic overlap above is not
tied to any particular diabatic or adiabatic representation. The orthonormal
electronic basis used to define the initial components can be selected
purely for physical convenience and independent of nuclear coordinates
$\boldsymbol{R}$,---that is, $\left\{ \left\vert \psi_{k}\left(\boldsymbol{R}\right)\right\rangle =\left\vert \psi_{k}\right\rangle \right\} ,$---without
affecting the primary mechanism under study. The exact-factorisation
analysis therefore identifies a representation-independent feature
of electron--nuclear composite systems: nuclear interference can
be viewed not only through amplitude transfer between electronic channels,
but also through the emergence of non-orthogonality among conditional
electronic states.

This perspective also clarifies the relation to previous studies in
which nuclear interference is analysed through how electronic properties
shape the potential governing nuclear motion. In the case of geometric-phase
effects, this influence is incorporated through the Born--Oppenheimer
potential energy surface augmented by the associated vector potential
\cite{Babikov2005044315,Yuan20181289}. In the exact-factorisation
framework, a closely related viewpoint appears in the form of the
exact time-dependent potential energy surface, which provides a fully
dynamical potential acting on the nuclei \cite{Curchod2016034103}.
In both cases, the emphasis lies on how electronic information is
transmitted to the nuclear subsystem through an effective potential
that determines the nuclear wave propagation. The present work situates
itself within this broader theme but adopts a complementary standpoint.
Rather than characterising nuclear interference through the structure
of an effective nuclear potential, we trace its emergence directly
to the dynamics of the electronic subsystem, identifying electronic
de-orthogonalisation as the mechanism by which electron--nuclear
correlations generate the interference contribution in the nuclear
density.

\section{Conclusion}

\label{conclu}

In summary, we have shown that starting from an initial superposition
of BO electronic states---where their mutual orthogonality ensures
the absence of interference in the nuclear density---subsequent non-adiabatic
dynamics can generate an interference contribution through the de-orthogonalisation
of the electronic factors within the exact factorisation framework.
While non-adiabatic coupling provides a natural mechanism for this
process, the more general underlying feature is the emergence of electronic
non-unitarity, which manifests itself through de-orthogonalisation
and gives rise to an interference term in the nuclear density. By
contrast, when the evolution is restricted to the adiabatic limit,
the electronic factors remain orthogonal and the nuclear density contains
no such interference term. Nuclear density interference therefore
emerges as a direct consequence of electron--nuclear correlation
effects inducing electronic subsystem de-orthogonalisation.

This perspective shifts the focus from interference within a single
subsystem to interference arising from the compositeness of the full
electron--nuclear state. The mechanism uncovered here is distinct
from conventional nuclear wave-packet interference within a single
BO channel and from interference that merely reflects electronic coherence
in a chosen representation. Instead, the interference term identified
in this work originates from correlation-induced de-orthogonalisation
between electronic components and manifests itself explicitly at the
level of the nuclear density.

Beyond its conceptual implications for correlated electron--nuclear
dynamics, this finding suggests that nuclear observables may serve
as complementary probes of the underlying non-adiabatic correlations.
More broadly, there is growing interest in engineering and controlling
non-unitary dynamics in quantum information science \cite{Northcote2025224112,Weidman2024102105}.
In particular, tailored electron--nuclear correlations have been
exploited to implement non-unitary logic operations in ion-trap platforms
\cite{Mourik2024040602}. The correlation-driven de-orthogonalisation
identified here, as a structural manifestation of non-unitary subsystem
evolution, may therefore provide a useful conceptual framework for
analysing and designing controlled non-unitary processes across diverse
composite quantum systems.

\appendix

\section{summary of the exact factorisation}

\label{EF-EOM}

\subsection{The exact equations of motion for factors of subsystems}

The exact factorisation method for coupled electrons and nuclei has
been formulated earlier in \cite{Abedi10123002,Abedi1222A530} with
numerous later applications, for example, Refs. \cite{Min2014263004,Requist16042108,Li2022113001,Arribas2024233201}.
Here we briefly summarise the main resulting equations of motion for
the ease of reference. The full electron-nuclear correlated wavefunction
is factorised exactly into a product between the nuclear and the electronic
factors, namely, 
\begin{equation}
\Psi\left(\boldsymbol{r},\boldsymbol{R},t\right)=\chi\left(\boldsymbol{R},t\right)\left\langle \boldsymbol{r}\left\vert \phi\left(t,\boldsymbol{R}\right)\right.\right\rangle .\label{EF-def}
\end{equation}
Here the nuclear factor $\chi\left(\boldsymbol{R},t\right)$ reproduces
the exact nuclear density and $N$-body nuclear current density while
the electronic factor $\left\vert \phi\left(t,\boldsymbol{R}\right)\right\rangle $
living in the electronic Hilbert space is parametrically dependent
on the nuclear coordinate $\boldsymbol{R}$ and subject to the partial
normalisation $\left\langle \phi\left(t,\boldsymbol{R}\right)\left\vert \phi\left(t,\boldsymbol{R}\right)\right.\right\rangle =1$
for each $\left(t,\boldsymbol{R}\right)$ .

The EOM satisfied by $\chi\left(\boldsymbol{R},t\right)$ and $\left\vert \phi\left(t,\boldsymbol{R}\right)\right\rangle $
are found bt substituting Eq. (\ref{EF-def}) into the TDSE for the
fully correlated electron-nuclear system. Explicitly, $\chi\left(\boldsymbol{R},t\right)$
satisfies an ordinary Schroedinger equation,
\begin{equation}
i\partial_{t}\chi\left(\boldsymbol{R},t\right)=\left\{ \mu\sum_{\nu}\frac{\left[\left(-i\boldsymbol{\nabla}_{\nu}\right)+\boldsymbol{A}_{\nu}\left[\phi\right]\left(\boldsymbol{R},t\right)\right]^{2}}{2M_{\nu}}+\varepsilon\left[\phi\right]\left(\boldsymbol{R},t\right)\right\} \chi\left(\boldsymbol{R},t\right).\label{EXF-EOM-nu}
\end{equation}
It evolves with an exact scalar potential energy surface (PES) $\varepsilon\left[\phi\right]\left(\boldsymbol{R},t\right)$
and exact vector potentials $\boldsymbol{A}_{\nu}\left[\phi\right]\left(\boldsymbol{R},t\right)$,
determined as functionals of the conditional electronic state $\left\vert \phi\left(t,\boldsymbol{R}\right)\right\rangle $.
The nuclear factor is thus also called the nuclear wavefunction in
the EF dictionary. The electronic factor is subject to a Schroedinger-like
equation,\begin{subequations}\label{e-EOM-set}
\begin{equation}
i\partial_{t}\left\vert \phi\left(t,\boldsymbol{R}\right)\right\rangle =\left[H^{BO}\left(\boldsymbol{R}\right)-\varepsilon^{A}\left[\phi\right]\left(\boldsymbol{R},t\right)+\mu V^{en}\left[\phi,\boldsymbol{\mathfrak{p}}\right]\left(\boldsymbol{R},t\right)\right]\left\vert \phi\left(t,\boldsymbol{R}\right)\right\rangle ,\label{e-EOM-full}
\end{equation}
where
\begin{equation}
V^{en}\left[\phi,\boldsymbol{\mathfrak{p}}\right]\left(\boldsymbol{R},t\right)=\left(U^{en}\left[\phi,\boldsymbol{\mathfrak{p}}\right]\left(\boldsymbol{R},t\right)-\varepsilon^{NA}\left[\phi\right]\left(\boldsymbol{R},t\right)\right).\label{Venmu}
\end{equation}
\end{subequations} which incorporates all non-adiabatic electron-nuclear
correlation effects. 

Explicitly, 
\begin{equation}
\boldsymbol{A}_{\nu}\left[\phi\right]\left(\boldsymbol{R},t\right)=\left\langle \phi\left(t,\boldsymbol{R}\right)\left\vert -i\boldsymbol{\nabla}_{\nu}\phi\left(t,\boldsymbol{R}\right)\right.\right\rangle ,\label{vecpt-def}
\end{equation}
 and 
\begin{equation}
\varepsilon\left[\phi\right]\left(\boldsymbol{R},t\right)=\varepsilon^{A}\left[\phi\right]\left(\boldsymbol{R},t\right)+\mu\varepsilon^{NA}\left[\phi\right]\left(\boldsymbol{R},t\right),\label{scapt-def}
\end{equation}
 with 
\begin{equation}
\varepsilon^{A}\left[\phi\right]\left(\boldsymbol{R},t\right)=\left\langle \phi\left(t,\boldsymbol{R}\right)\left\vert \left(H^{BO}\left(\boldsymbol{R}\right)-i\frac{\partial}{\partial t}\right)\right\vert \phi\left(t,\boldsymbol{R}\right)\right\rangle ,\label{scapt-gd}
\end{equation}
 and 
\begin{equation}
\varepsilon^{NA}\left[\phi\right]\left(\boldsymbol{R},t\right)=\left\langle \phi\left(t,\boldsymbol{R}\right)\right\vert U_{K}^{en}\left[\phi\right]\left(\boldsymbol{R},t\right)\left\vert \phi\left(t,\boldsymbol{R}\right)\right\rangle .\label{scapt-gi}
\end{equation}
The electron-nuclear correlation operator is given by 
\begin{equation}
U^{en}\left[\phi,\boldsymbol{\mathfrak{p}}\right]\left(\boldsymbol{R},t\right)=U_{K}^{en}\left[\phi\right]\left(\boldsymbol{R},t\right)+U_{Q}^{en}\left[\phi,\boldsymbol{\mathfrak{p}}\right]\left(\boldsymbol{R},t\right)\label{Uen-def}
\end{equation}
with 
\begin{equation}
U_{K}^{en}\left[\phi\right]\left(\boldsymbol{R},t\right)=\sum_{\nu}\frac{\left[\left(-i\boldsymbol{\nabla}_{\nu}\right)-\boldsymbol{A}_{\nu}\left[\phi\right]\left(\boldsymbol{R},t\right)\right]^{2}}{2M_{\nu}},\label{EXF-enucor-K}
\end{equation}
\begin{equation}
U_{Q}^{en}\left[\phi,\boldsymbol{\mathfrak{p}}\right]\left(\boldsymbol{R},t\right)=\sum_{\nu}\boldsymbol{\mathfrak{p}}_{\nu}\left[\phi,\chi\right]\left(\boldsymbol{R},t\right)\cdot\frac{\left[-i\boldsymbol{\nabla}_{\nu}-\boldsymbol{A}_{\nu}\left[\phi\right]\left(\boldsymbol{R},t\right)\right]}{M_{\nu}},\label{Uen_Q}
\end{equation}
in which $\boldsymbol{\mathfrak{p}}\left(\boldsymbol{R},t\right)=\left\{ \boldsymbol{\mathfrak{p}}_{\nu}\left[\phi,\chi\right]\left(\boldsymbol{R},t\right)\right\} $
given by

\begin{equation}
\boldsymbol{\mathfrak{p}}_{\nu}\left[\phi,\chi\right]\left(\boldsymbol{R},t\right)=\left[-i\frac{\boldsymbol{\nabla}_{\nu}\chi\left(\boldsymbol{R},t\right)}{\chi\left(\boldsymbol{R},t\right)}+\boldsymbol{A}_{\nu}\left[\phi\right]\left(\boldsymbol{R},t\right)\right],\label{pzero-q}
\end{equation}
is the so-called nuclear momentum function. It is complex in general
and its imaginary part, $\text{Im}\left\{ \boldsymbol{\mathfrak{p}}_{\nu}\right\} =-i\boldsymbol{\nabla}_{\nu}\left\vert \chi\right\vert /\left\vert \chi\right\vert $,
is often called the nuclear quantum momentum \cite{Min15073001,Arribas2024233201}.
Its real part divided by the nuclear mass $M_{\nu}$, on the other
hand, is the nuclear velocity field, $\text{Re}\left\{ \boldsymbol{\mathfrak{p}}_{\nu}\right\} /M_{\nu}=\left(\boldsymbol{\nabla}_{\nu}\text{arg}\left(\chi\right)+\boldsymbol{A}_{\nu}\right)/M_{\nu}=\boldsymbol{J}_{\nu}/\left(\left\vert \chi\right\vert ^{2}M_{\nu}\right)$.
$\boldsymbol{J}_{\nu}$ is the gauge-invariant nuclear current density.

\subsection{The adiabtic approximation}

\label{EF-EOM-ADA}

By applying Born-Huang-like expansion to the EF electronic factor,
\begin{equation}
\left\vert \phi\left(t,\boldsymbol{R}\right)\right\rangle =\sum_{k}C_{k}\left(t,\boldsymbol{R}\right)\left\vert \varphi_{k}\left(\boldsymbol{R}\right)\right\rangle ,\label{BH-EF-ce1}
\end{equation}
the $k$th BH coefficient $\chi_{k}\left(\boldsymbol{R},t\right)$
are related to the EF nuclear and electronic factors by 
\begin{equation}
\chi_{k}\left(\boldsymbol{R},t\right)=\chi\left(\boldsymbol{R},t\right)C_{k}\left(t,\boldsymbol{R}\right).\label{BH-EF-r1}
\end{equation}
The adiabatic EOM for $\left\vert \chi_{k}^{ad}\left(t\right)\right\rangle $
given by Eq. (\ref{ad-HBk-eq}) can be derived from the EF EOMs Eqs.
(\ref{EXF-EOM-nu}) and (\ref{e-EOM-set}) via the defining relation
Eq. (\ref{BH-EF-r1}). From Eq. (\ref{e-EOM-set}) for the electronic
factor, one can find the corresponding EOMs for $\left\{ C_{k}\left(t,\boldsymbol{R}\right)\right\} $
(appearing in the BH-like expansion Eq. (\ref{BH-EF-ce1})) involving
nuclear-coordinate derivative of both $\left\{ C_{k}\left(t,\boldsymbol{R}\right)\right\} $
and the BO electronic states $\left\{ \left\vert \varphi_{k}\left(\boldsymbol{R}\right)\right\rangle \right\} $.
As the adiabtic approximation, we remove terms that induce inter-BO-state
transitions in the EOM for $\partial_{t}C_{k}\left(t,\boldsymbol{R}\right)$
and that leads to $\left\vert \phi\left(t,\boldsymbol{R}\right)\right\rangle \approx C_{k}\left(t,\boldsymbol{R}\right)\left\vert \varphi_{k}\left(\boldsymbol{R}\right)\right\rangle $
where by $\left\langle \phi\left(t,\boldsymbol{R}\right)\left\vert \phi\left(t,\boldsymbol{R}\right)\right.\right\rangle =1$
one finds $C_{k}\left(t,\boldsymbol{R}\right)=e^{i\theta_{k}\left(t,\boldsymbol{R}\right)}$
is a phase factor. Maintaining still the coordinate derivatives of
$C_{k}\left(t,\boldsymbol{R}\right)$, then combining $\partial_{t}C_{k}\left(t,\boldsymbol{R}\right)$
with Eq. (\ref{EXF-EOM-nu}) for $\chi\left(\boldsymbol{R},t\right)$
using the chain rule, $\partial_{t}\chi_{k}\left(\boldsymbol{R},t\right)=\left[\partial_{t}C_{k}\left(t,\boldsymbol{R}\right)\right]\chi\left(\boldsymbol{R},t\right)+C_{k}\left(t,\boldsymbol{R}\right)\partial_{t}\chi\left(\boldsymbol{R},t\right)$,
we then recover $i\partial_{t}\chi_{k}\left(\boldsymbol{R},t\right)=\left\{ \mu\sum_{\nu}\left[\left(-i\boldsymbol{\nabla}_{\nu}+\boldsymbol{A}_{\nu}\left[C_{k}\varphi_{k}\right]\left(\boldsymbol{R},t\right)\right)\right]^{2}/\left(2M_{\nu}\right)+\varepsilon_{k}\left(\boldsymbol{R}\right)+\mu\varepsilon^{NA}\left[C_{k}\varphi_{k}\right]\left(\boldsymbol{R},t\right)\right\} \chi_{k}\left(\boldsymbol{R},t\right)$
as Eq. (\ref{ad-HBk-eq}). Comparing with the usual single-surface
adiabatic ansatz $\Psi\left(\boldsymbol{r},\boldsymbol{R},t\right)=\chi_{k}\left(\boldsymbol{R},t\right)\left\langle \boldsymbol{r}\left\vert \varphi_{k}\left(\boldsymbol{R}\right)\right.\right\rangle $,
the EF derivation shows that the phase factor $C_{k}\left(t,\boldsymbol{R}\right)=e^{i\theta_{k}\left(t,\boldsymbol{R}\right)}$
corresponds to the gauge freedom of the factorisation. The EOM derived
above therefore reproduces Eq. (\ref{ad-HBk-eq}) up to the corresponding
gauge transformation, consistent with the well-established gauge covariance
of the EF formalism \cite{Abedi10123002,Abedi1222A530}.

\subsection{Perturbation by the non-adiabatic correlation }

The structure of Eq. (\ref{e-EOM-set}) furnishes a perturbation expansion
of the electronic factor (see details also in \cite{Tu2025ARXIV251102004}),
\begin{equation}
\left\vert \phi\left(t,\boldsymbol{R}\right)\right\rangle =\left\vert \phi^{\left(0\right)}\left(t,\boldsymbol{R}\right)\right\rangle +\mu\left\vert \phi^{\left(1\right)}\left(t,\boldsymbol{R}\right)\right\rangle +\mathcal{O}\left(\mu^{2}\right)\label{eF-pert-expn}
\end{equation}
where the unperturbed state evolution follows 
\begin{equation}
i\partial_{t}\left\vert \phi^{\left(0\right)}\left(t,\boldsymbol{R}\right)\right\rangle =\left(H^{BO}\left(\boldsymbol{R}\right)-\varepsilon^{A}\left[\phi^{\left(0\right)}\right]\left(\boldsymbol{R},t\right)\right)\left\vert \phi^{\left(0\right)}\left(t,\boldsymbol{R}\right)\right\rangle \label{eF-pert-0}
\end{equation}
The non-adiabatic effects can now be accounted for by including the
first-order correction $\left\vert \phi^{\left(1\right)}\left(t,\boldsymbol{R}\right)\right\rangle $
in Eq. (\ref{eF-pert-expn}). Explicitly this is given by
\begin{equation}
\left\vert \phi^{\left(1\right)}\left(t,\boldsymbol{R}\right)\right\rangle =-i\int_{0}^{t}\text{d}t^{\prime}U^{BO}\left(t-t^{\prime},\boldsymbol{R}\right)\left\vert \xi^{\left(0\right)}\left(t^{\prime},\boldsymbol{R}\right)\right\rangle ,\label{1st-stat1}
\end{equation}
where $U^{BO}\left(t-t^{\prime},\boldsymbol{R}\right)=e^{-i\left(t-t^{\prime}\right)H^{BO}\left(\boldsymbol{R}\right)}$
is the evolution operator governed by the BO Hamiltonian for the zeroth-order
state while 
\begin{equation}
\left\vert \xi^{\left(0\right)}\left(t^{\prime},\boldsymbol{R}\right)\right\rangle =V^{en}\left[\phi^{\left(0\right)},\chi\right]\left(\boldsymbol{R},t^{\prime}\right)\left\vert \phi^{\left(0\right)}\left(t^{\prime},\boldsymbol{R}\right)\right\rangle .\label{1st-stat2}
\end{equation}

\section*{Acknowledgement}

E.K.U.G. acknowledges support from fondation de l'\'{E}cole Polytechnique,
Palaiseau, France, within the Gaspard Monge Programme. This work was
supported by the European Research Council (ERC-2024-SyG- 101167294;
UnMySt) and by the Cluster of Excellence 'CUI: Advanced Imaging of
Matter' of the Deutsche Forschungsgemeinschaft (DFG) - EXC 2056 -
project ID 39071599. We acknowledge support from the Max Planck-New
York City Center for Non-Equilibrium Quantum Phenomena. 

\bibliographystyle{myunsrt} 
\bibliography{/Users/matissetour/Documents/Forschung/filetex/centralbib/refs_bibexd-1.bib} 

\begin{thebibliography}{10}
\bibitem{FootNoteInitFullCor}
If we have the initial superposition given by
  $\Psi\left(\boldsymbol{r},\boldsymbol{R},t=0\right)=\sum_{k}c_{k}\Psi_{k}^{M}\left(\boldsymbol{r},\boldsymbol{R}\right)$,
  where $\Psi_{k}^{M}\left(\boldsymbol{r},\boldsymbol{R}\right)$ is the $k$th
  eigentate of the fully correlated system with eigenenergy $E_{k}$ independent
  of $\left(\boldsymbol{r},\boldsymbol{R}\right)$, then the general time
  evolution Eq. (\ref{superpos-pr}) still holds but with
  $\Psi_{k}\left(\boldsymbol{r},\boldsymbol{R},t=0\right)=\Psi_{k}^{M}\left(\boldsymbol{r},\boldsymbol{R}\right)$.
  Subsequently, the nuclear density is still given formallly by Eq.
  (\ref{nud-inf-itp}). Nevertheless, this will result in
  $n_{jk}\left(\boldsymbol{R},t\right)=\exp\left\{
  -it\left(E_{k}-E_{j}\right)\right\} \bar{n}_{jk}\left(\boldsymbol{R}\right)$
  in which the spatial dependence of the nuclear density,
  $\bar{n}_{jk}\left(\boldsymbol{R}\right)=\left[\int\text{d}\boldsymbol{r}\Psi_{j}^{M*}\left(\boldsymbol{r},\boldsymbol{R}\right)\Psi_{k}^{M}\left(\boldsymbol{r},\boldsymbol{R}\right)\right]$,
  becomes fully constant in time and factorised from its time dependence. Such
  initial states are therefore not suitable for examining the time-dependent
  emergence of nuclear interference from an initially interference-free
  configuration, which constitutes the primary objective of this study.
\end{thebibliography}
\end{document}